\documentclass[pra,twocolumn,amsmath,amssymb,aps,longbibliography,showpacs]{revtex4-1}
\usepackage{graphicx}% Include figure files
\usepackage{dcolumn}% Align table columns on decimal point
\usepackage{bm}% bold math
\usepackage{CJK}
\usepackage{subfigure}
\usepackage{color}
\usepackage{epstopdf}

\begin{document}

%\preprint{APS/123-QED}

\title{Universality in nonlinear  passage through the miscible-immiscible phase transition in two component Bose-Einstein condensates}

\author{Xunda Jiang$^{1}$, Yikai Ji$^{1}$, Bin Liu$^{1}$, Feng Li$^{1}$, Xizhou Qin$^{*,1}$, Yongyao Li$^{1}$ and Chaohong Lee$^{2}$}

\address{$^1$School of Physics and Optoelectronic Engineering, Foshan University, Foshan 528000, China}
\address{$^2$College of Physics and Optoelectronic Engineering, Shenzhen University, Shenzhen 518060, China}
\email{XizhouQin@fosu.edu.cn}

\date{\today}% It is always \today, today,
             %  but any date may be explicitly specified

\begin{abstract}
In this study, we investigate the formation of domain defects and the universal critical real-time dynamics in a two-component Bose-Einstein condensate with nonlinear quenching across the miscible-immiscible phase transition.
By analyzing the Bogoliubov excitations, we obtain the power-law relations among the defect density, the phase transition delay and the quench time near the phase transition.
Moreover, by simulating the real-time dynamics across the miscible-immiscible phase transition, we clearly show the formation of domain defects and the delay of the phase transition.
Furthermore, we find that the domain defects are suppressed by large nonlinear coefficients and long quench times.
To accurately characterize the domain defects, we quantify the defect excitations using the correlation length and the domain number.
In addition, by combining the power-law relations between the phase transition delay and the quench time, we extract the critical exponents for different nonlinear coefficients.
Our study not only confirms that the critical exponents do not sensitively depend on the nonlinear quenches but also provides a dynamic path toward the suppression of nonadiabatic excitation.
\end{abstract}

%\pacs{Valid PACS appear here}% PACS, the Physics and Astronomy
                             % Classification Scheme.

%\keywords{Suggested keywords}%Use showkeys class option if keyword
                              %display desired
% \noindent{\it Keywords}: defects formation, nonlinear quenching, universal dynamics, critical exponents, Bose-Einstein condensates
\maketitle

%\tableofcontents

\section{INTRODUCTION}

Continuous quantum phase transitions, which are associated with spontaneous symmetry breaking~\cite{Sachdev2011,Morikawa1995,Kibble1980}, have been extensively studied for many years in various systems.
When the system crosses the phase transition, the adiabatic theorem fails, and non-adiabatic topological defects, such as solitons~\cite{Zurek2010,Witkowska2011,Zurek2009}, domains~\cite{Kibble1976,Lee2009,Davis2011,Davis2012,Xu2016,Swislocki2013, Hofmann2014,Wu2017,Navon2015,Ye2018}, or vortices~\cite{Anderson2008,Su2013,Wu2016}, are inevitably generated.
The formation of the defects and their universal dynamics can be described by a paradigmatic theory called the Kibble-Zurek mechanism (KZM)~\cite{Kibble1976,Kibble1980,Zurek1985,Zurek1996,JDziarmaga2000,Polkovnikov2011,Bloch2008}, %which is a paradigmatic theory used to describe the universality of defects formation.
The KZM has been extensively studied in a number of different systems, such as condensed matter systems~\cite{Ruutu1996,Bauerle1996,Monaco2009}, quantum many-body systems~\cite{ Keesling2019,Harris2018,Eisert2015,Polkovnikov2011}, and ultracold atomic gases~\cite{Anderson2008,Zurek2009,Lee2009,Qiu2020,Damski2010,Witkowska2011,Davis2011,Davis2012,Navon2015,Lamporesi2013,Anquez2016,Clark2016,Feng2018,Swislocki2013,Hofmann2014,Wu2017,Xu2016,Ye2018,Jiang2019,Jiang2021}.

In recent years, adiabatic quantum computation has become a very fascinating topic for simulating certain physical problems~\cite{Albash2018,Gyongyosi2019,Ladd2010,Bloch2008}.
The key for adiabatic quantum computation is to prepare the adiabatic ground state and then adiabatically quench the Hamiltonian to a nontrivial regime~\cite{Albash2018}.
To encode nontrivial information, quenching the system across the continuous phase transitions between the two phases is sometimes necessary. However, it is quite difficult to adiabatically maintain a sufficiently slow evolution so that the system will always be excited. For finite-size quantum many-body systems, conventional methods that suppress non-adiabatic excitations sacrifice the computational time by allowing the system to smoothly cross the critical point without excitation, while such a proposal is not realistic for adiabatic quantum computations~\cite{Polkovnikov2008}.
Moreover, several recent works have investigated the acceleration of evolution to reduce nonadiabatic defects.
These schemes are similar to shortcuts to adiabaticity~\cite{Odelin2019}, which require complicated mathematical calculations to design a shortcut path to the final desired state.
On the other hand, nonlinear quenching across the critical point is also an approach to reduce nonadiabatic excitations in recent studies~\cite{Barankov2008,Sen2008,Mondal2009}, and such an approach is experimentally accessible and intuitive to understand.
However, those relative works mainly focus on quantum many-body systems, with a slight focus on mean-field quantum systems.

In this paper, we investigate the formation of defects in two-component BECs in nonlinear quenching across a miscible-immiscible phase transition.
From the Bogoliubov excitation spectrum, we extract the dynamic critical exponent and the static correlation length critical exponent.
According to the KZM, the universal critical exponents depend only on the dimension of the system, the symmetry and the type of interaction, which are independent of the quench type.
In contrast, nonlinear quenching affects the power-law relation among the defect density, phase transition delay, and quench time.
From the numerical simulation, we clearly show the effect of nonlinear quenching on the formation of the domain defects.
By calculating the phase transition delay and defect density, we confirm that the critical exponents do not sensitively depend on the nonlinear quenches of the miscible-immiscible phase transition.
Finally, we show that nonlinear quenching with large power coefficients can suppress the nonadiabatic excitation of the system and provide a path toward adiabatic evolution.

The paper is organized as follows.
In Sec.~\uppercase\expandafter{\romannumeral2}, we describe the physical model and the key idea of KZM.
In Sec.~\uppercase\expandafter{\romannumeral3}, we analyze the Bogoliubov excitations and analytically extract the critical exponents from the fast-growing modes.
In Sec.~\uppercase\expandafter{\romannumeral4}, we present real-time nonequilibrium dynamics and extract critical exponents from the phase transition delays and divergence length scales.
Finally, we provide a brief summary and discussion in Sec.~\uppercase\expandafter{\romannumeral5}.

\section{Model}

We consider two-component BECs confined in a one-dimensional homogeneous trap, which can be described by the Gross-Pitaevskii equations (GPEs),
\begin{equation}\label{TGPE1}
i\hbar \frac{{\partial {\psi _j}}}{{\partial t}} = \left[ { - \frac{{{\hbar ^2}}}{{2{m_j}}}\frac{{{\partial ^2}}}{{\partial {x^2}}} + {g_{jj}}{{\left| {{\psi _j}} \right|}^2} + {g_{12}}{{\left| {{\psi _{3 - j}}} \right|}^2}} \right]{\psi _j},
\end{equation}
where $g_{jj}=2\hbar^{2}a_{jj}/(ma_{\bot}) \left( j=1,2\right)$ and $g_{12}=2\hbar^{2}a_{12}(m_{1}+m_{2})/(m_{1}m_{2}a_{\bot})$ represent the strength of the intra-component interaction and inter-component interaction, respectively; here, $a_{\bot}$ represents the length scale in the transverse direction. The subscript $j$ denotes different components, and $\psi_{j}$ represents the
wave function for different BECs, which are normalized as $\int {{{\left| {{\psi _j}} \right|}^2}dx}  = {N_j}$.
The competition between the intra-component interaction $g_{jj}$ and inter-component interactions $g_{12}$ results in two distinct phases~\cite{Timmermans1998,Ao1998,Trippenbach2000,Davis2011,Davis2012,Jiang2021}, i.e., the immiscible phase and the miscible phase, and the phase transition
between them occurs when $g_{12}^{2}=g_{11}g_{22}$. For strong intra-component interactions,
i.e. $g_{12}^{2}<g_{11}g_{22}$, the two-component BECs prefer to coexist everywhere, corresponding
to lower energies for the whole system. For a strong inter-component interaction, i.e., $g_{12}^{2}>g_{11}g_{22}$, however, two-component BECs prefer to be separated from each other.

To investigate the universal dynamics of the system, we introduce a dimensionless distance
\begin{equation}\label{Dim_g_1}
\epsilon \left( t \right) = \left| {{g_{22}}\left( t \right) - g_{22}^c} \right|/g_{22}^c.
\end{equation}
Then, we quench the interaction strengths $g_{22}$ across the critical point $g_{22}^c$, giving
\begin{equation}\label{Non_Quen_1}
 {g_{22}}\left( t \right) = g_{22}^c\left( {1 + \mathrm{sgn}\left( t \right){{\left| {\frac{t}{{{\tau _Q}}}} \right|}^r}} \right).
\end{equation}
Here, $\tau _Q$ is the quench time, which characterizes how fast the system crosses the critical point, $r$ denotes the power exponent, and sgn is the signum function. From the KZ argument, the system becomes adiabatic near the freeze time $\hat{t}$, and the related $\hat{\epsilon}$ is given by
\begin{equation}\label{Free_time}
 \hat{\epsilon} \sim \left(r/\tau_{Q}\right)^{\frac{r}{1+rvz }}, \quad \hat t = {\left( {r{\tau _Q}^{rvz}} \right)^{\frac{1}{{rvz + 1}}}},
\end{equation}
the correlation length and the density of the excitations that freeze at $\hat{t}$ are
\begin{equation}\label{Correl_len_1}
\hat{\xi} \sim \tau_{Q}^{\frac{rv}{1+rvz}}, \quad  n_{ex} \simeq \hat{\xi}^{-d} \sim \tau_{Q}^{-\frac{rvd}{1+rvz}},
\end{equation}
where $d$ is the number of space dimensions.

\section{BOGOLIUBOV EXCITATION AND CRITICAL EXPONENTS}

In this section, we retrospectively describe the Bogoliubov excitations near the critical point of the phase separation and extract the critical exponents~\cite{Timmermans1998,Ao1998,Trippenbach2000,Jiang2019}. By applying the Bogoliubov analysis, the excitation modes beyond the ground state can be obtained. Regarding the miscible phase, the nonlinear Schr$\ddot{o}$dinger equation has an obvious homogenous solution $\rho_{j}=|\phi_{j}|^{2}=N_{j}/L$, with $\phi_{j}$ being the ground state wave-function of the $j$-th component, and $L$ being the length of the system.
The chemical potentials are given as $\mu_{1}=g_{11}\rho_{1}+g_{12}\rho_{2}$ and $\mu_{2}=g_{22}\rho_{2}+g_{12}\rho_{1}$.
To obtain the Bogoliubov excitation spectrum, the perturbed ground state is considered as$($setting $\hbar=m=1)$
\begin{equation}
 \psi_{j}\left(x,t \right)=[\phi_{j}+\delta \phi_{j}\left(x,t\right)]e^{-i\mu_{j}t},   \label{Bogoliubov form1}
 \end{equation}
where the perturbations $\delta\phi_{1,2}$ can be written as
\begin{equation}
\left( {\begin{array}{*{20}{c}}
{\delta {\phi _1}}\\
{\delta {\phi _2}}
\end{array}} \right) = \left( {\begin{array}{*{20}{c}}
{{u_{1,q}}}\\
{{u_{2,q}}}
\end{array}} \right){e^{iqx - i\omega t}} + \left( {\begin{array}{*{20}{c}}
{{v_{1,q}^{*}}}\\
{{v_{2,q}^{*}}}
\end{array}} \right){e^{iqx + i\omega t}}.  \label{fluctuation form1}
\end{equation}
Here, $q$ and $\omega$ represent the excitation quasimomentum and excitation frequency, respectively. $u_{j,q}$ and $v_{j,q}$, $(j=1,2)$ are the Bogoliubov amplitudes. By comparing the coefficients for the terms $e^{iqx-i\omega t}$ and $e^{iqx+i\omega t}$, the BdG equations can be obtained
\begin{equation}
\mathcal{M}(q) \mathbf{u}_{q}=\omega  \mathbf{u}_{q},
\end{equation}
\begin{equation}
\mathcal{M}(q) = \left( {\begin{array}{*{20}{c}}
{\hat{h}_{1}} &{g_{12}\phi_{1}\phi_{2}} &{g_{11}\phi_{1}^2} &{g_{12}\phi_{1}\phi_{2}}\\
{g_{12}\phi_{1}\phi_{2}} &{\hat{h}_{2}}&{g_{12}\phi_{1}\phi_{2}} &{g_{22}\phi_{2}^2}\\
{-g_{11}\phi_{1}^2} &{-g_{12}\phi_{1}\phi_{2}}&{-\hat{h}_{1}} &{-g_{12}\phi_{1}\phi_{2}}\\
{-g_{12}\phi_{1}\phi_{2}} &{-g_{22}\phi_{2}^2} &{-g_{12}\phi_{1}\phi_{2}} &{-\hat{h}_{2}}
\end{array}} \right),
\end{equation}
where $\mathbf{u}_{q} =\left(u_{1,q},u_{2,q},v_{1,q},v_{2,q}\right)^{T}$ and $\hat{h}_{j}=\frac{q^{2}}{2} +\sum_{k}g_{jk}|\phi_{k}|^2+g_{jj}|\phi_{j}|^2-\mu_{j}$.
For the miscible phase, the two BECs are assumed to have the same particle number, i.e., $N_{1}=N_{2}=N/2$.
The excitation spectrum can be obtained by diagonalizing the matrix $\mathcal{M}(q) $ for each $q$
\begin{equation}
\omega_{\pm}^{2}=\epsilon_{0}\left(\epsilon_{0}+2\eta_{\pm} \right),    \label{excitation}
\end{equation}
where $\epsilon_{0}=q^2/2$, and
\begin{equation}
\eta_{\pm}=\frac{\rho}{4}\left(g_{11}+g_{22}\pm \sqrt{\left( g_{11}-g_{22}  \right)^{2}+4g_{12}^{2}}   \right).
\end{equation}
We then derive the two critical exponents from the Bogoliubov excitations~\cite{Timmermans1998,Jiang2019}. In the long-wavelength limit of $q\to 0$, the sound velocity becomes
\begin{equation}
c_ {\pm} = \left. {\frac{{\partial {\omega _ {\pm}}}}{{\partial q}}} \right|_{q \to 0}=\sqrt{\eta_{\pm}} \propto\left|\epsilon\right|^{1/2}.
\end{equation}

In the immiscible phase, i.e., $g_{12}^{2}>g_{11}g_{22}$, $c_{-}$ becomes a complex value, leading to dynamical instabilities in the following evolution. The fast unstable mode will grow exponentially, which
is given by minimizing $\omega_{-}^{2}$ with respect to $q$, i.e., $\partial  \omega_{-}^{2}/\partial q=0$,
giving ${q_f} = \sqrt 2 \left| {{c_ - }} \right|$, and its corresponding energy scale is $E_{q}={\left| {{c_ - }} \right|^2}$~\cite{Jiang2019,Timmermans1998}. From the dimensional analysis, we expect that the length scale $\xi$ of the defect modes generated in the immiscible phase has the form given by
\begin{equation}
\xi  = 2\pi /{q_f} = \sqrt {\rm{2}} \pi /\left| {{c_ - }} \right| \propto {\left|\epsilon\right|^{ - 1/2}},    \label{length_scale1}
\end{equation}
In addition, we estimate that the time scale for the generation of defects has
\begin{equation}
\tau  = 1/{E_q} = 1/{\left| {{c_ - }} \right|^2} \propto {\left|\epsilon\right| ^{ - 1}}.
\label{time_scale1}
\end{equation}
Comparing Eq.~\ref{length_scale1} and Eq.~\ref{time_scale1} with the KZ argument, we extract the critical exponents $\left(v=1/2, z=2\right)$.

\begin{figure}[htbp]
\centering
\includegraphics[scale=0.42]{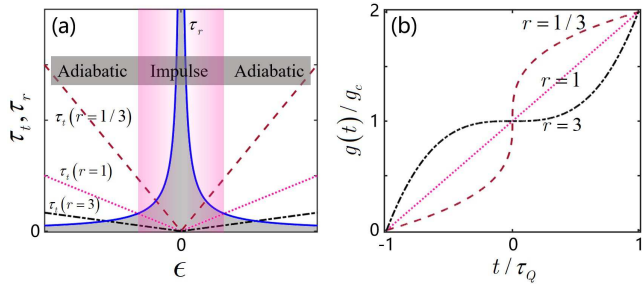}
\caption{(Color online) (a) The relationship between the relaxation time $\tau_{r}$, transition time $\tau_{t}$ and dimensionless distance $\epsilon$. (b) The time-dependent strength of the interaction for different power exponents $r$.}
\label{Non_KZ_diagram_1}
\end{figure}

\section{Numerical simulation}
In this section, we numerically investigate the non-equilibrium dynamics of the system by quenching the interaction $g_{22}(t)$ across the critical point. In our quenched approach, we choose various ways of traversing the critical point according to Eq.\ref{Non_Quen_1} with different power coefficients $r$.
From numerical simulations, we can visibly show the non-adiabatic defects and other critical behaviors, such as phase transition delays, and then numerically extract the critical exponents from the non-equilibrium dynamics. To perform the quenched dynamics, we first prepare the initial state in the miscible phase, where the densities of the two component BECs share all the spatial space. We then increase the intra-species interaction $g_{22}(t)$ according to Eq.\ref{Non_Quen_1} for various quench times $\tau_{Q}$ and power coefficients $r$. Due to the vanishing of the energy gap at the critical point, the relaxation time diverges near the critical point, resulting in instantaneous states that cannot adiabatically follow the change in the Hamiltonian. According to the adiabatic-impulse-adiabatic approximation, the instantaneous state will be frozen in the impulse region separated by $\pm \hat(t)$. The frozen state indicates that it nearly stops evolving in the impulse region. When the quench parameter is out of the impulse region, the state is restarted for evolution. At this time, the states frozen at $-\hat{t}$ are no longer the eigenstates of the Hamiltonian at $+\hat{t}$. Therefore, non-adiabatic defects are inevitably generated, and the defect density obeys a KZ scaling for different quenching times.
\begin{figure}[htbp]
\centering
\includegraphics[scale=0.42]{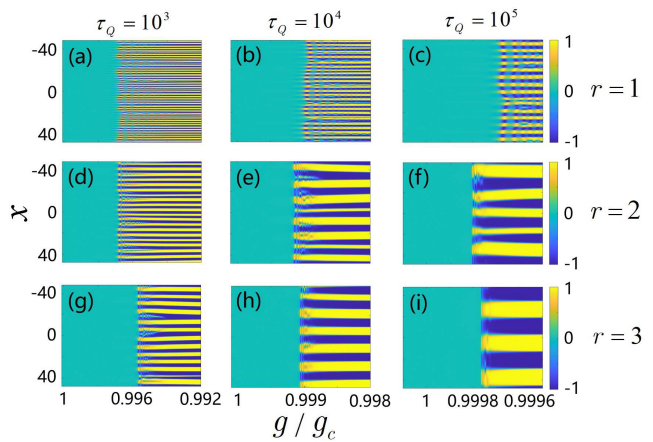}
\caption{(Color online) Examples of the domain formation in quenched dynamics for different quench times
$\tau_Q$ and power coefficients $r$. The first row to the third row correspond to $r=1,2,3$, and the three columns correspond to different quench times $\tau_{Q}$, as displayed in the figure. The other parameters are chosen as $N=2\times 10^{6}$, $L=96$, and $g_{11}=g_{12}=0.5$.}
\label{Fig_Wave_Func_1}
\end{figure}

 In Fig.\ref{Fig_Wave_Func_1}, we typically show the domain formation in the density difference $J\left(x\right)$ of the two-component BECs, where $J\left(x\right)$ is defined as
\begin{equation}\label{den_diff_1}
   J\left(x\right)=\frac{n_{1}\left(x\right)-n_{2}\left(x\right)}{n_{1}\left(x\right)+n_{2}\left(x\right)}, \quad n_{j}\left(x\right)=\left| \psi_{j} \left(x\right)\right|^{2}.
 \end{equation}
For smaller $\tau_{Q}$, the parameter in the Hamiltonian changes more quickly. Hence, the system becomes more non-adiabatic as it crosses the critical point, which results in more defect generations in the subsequent evolution (see Fig. \ref{Fig_Wave_Func_1}(a,d,g)). In our system, the non-adiabatic defect corresponds to the domain wall shown in Fig. \ref{Fig_Wave_Func_1}. For larger $\tau_{Q}$, the number of non-adiabatic defects will decrease, and the system will evolve more adiabatically (see Fig.\ref{Fig_Wave_Func_1} (c, f, i). It is expected that when $\tau_{Q}\to\infty$, the non-adiabatic defect vanishes, and the nonequilibrium dynamics return to the equilibrium case.

\begin{figure}[htbp]
\centering
\includegraphics[scale=0.42]{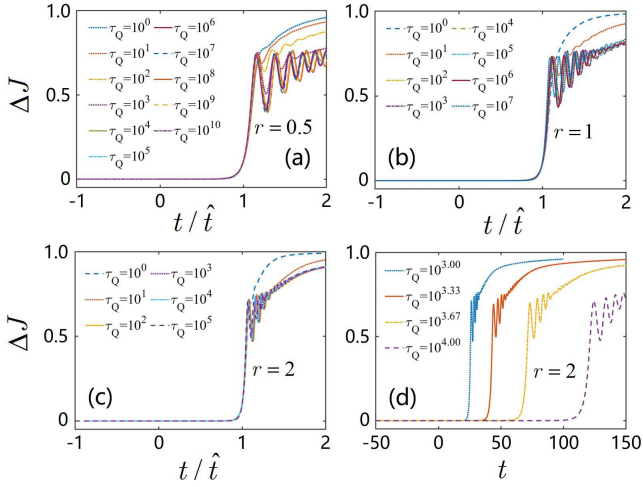}
\caption{(Color online) The spatial fluctuation of the local spin polarization $\Delta {J}$ in quantum critical dynamics for different power coefficients, $r=0.5$(a), $r=1$(b), and $r=2$(c,d).}
\label{Fig_Order_Para_1}
\end{figure}

To quantitatively describe the process of the evolution, we define the spatial fluctuations of the local spin
polarization as
\begin{equation}
\Delta {J} = \sqrt {\frac{1}{L}\int {J^2\left( x \right)dx}
- {{\left[ {\frac{1}{L}\int {{J}\left( x \right)dx} } \right]}^2}}.
\end{equation}
Here, $L$ is the length of the system, and $\Delta {J}$ plays the role of an order parameter that describes where the phase transition occurs. In Fig.\ref{Fig_Order_Para_1}, we clearly show the $\Delta {J}$
vary over time for different $r$.
According to KZM, the instantaneous state freezes at $-\hat{t}$ and remains unchanged until the system
passes over $+\hat{t}$. During the first adiabatic region and the impulse region, the instantaneous state retains most of the information in the miscible phase. In this case, the spatial fluctuation of the local spin polarization $\Delta {J}$ remains zero.
However, when the system crosses $+\hat{t}$, the instantaneous state starts to evolve again, but the state at this moment is no longer an eigenstate of the Hamiltonian. Therefore, the instantaneous state that freezes at $-\hat{t}$ will experience a non-adiabatic process to a symmetry breaking state, resulting in non-adiabatic defects and non-zero $\Delta {J}$ occurring after $+\hat{t}$. Typical examples of $\Delta {J}$ in this process are shown in Fig.\ref{Fig_Order_Para_1} (d). We determined
the freezing time $\hat{t}$ when $\Delta {J}$ reaches a small nonzero value $\delta_{J}$. In our simulation, $\delta_{J}$ is chosen as $0.05$. Based on our calculations, a similar conclusion can be drawn for other $\delta_{J}$ ranges from $0.05$ to $0.5$. In Fig. \ref{Fig_Order_Para_1}(a,b,c), we show the
$\delta_{J}$ varies with the rescaled time $t/\hat{t}$ for different quench times $\tau_{Q}$ and coefficients $r$, and the collapse of $\delta_{J}$ for different $\tau_{Q}$ indicates that there exists a scaling law between $\hat{t}$ and quench time $\tau_{Q}$.
\begin{figure}[htbp]
\centering
\includegraphics[scale=0.35]{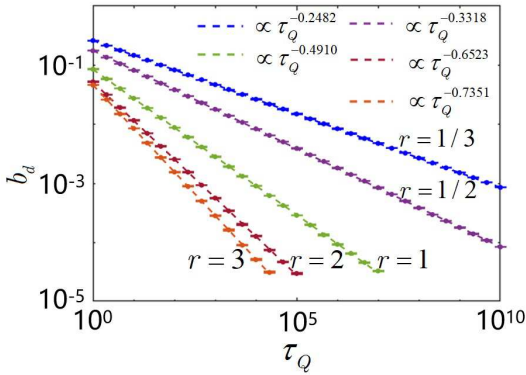}
\caption{(Color online) The temporal universal scaling of the phase transition delay $b_{d}$
with respect to the quench time $\tau_{Q}$ for different $r$. }
\label{Fig_Delay1_vs_Tauq}
\end{figure}

\begin{figure}[htbp]
\centering
\includegraphics[scale=0.44]{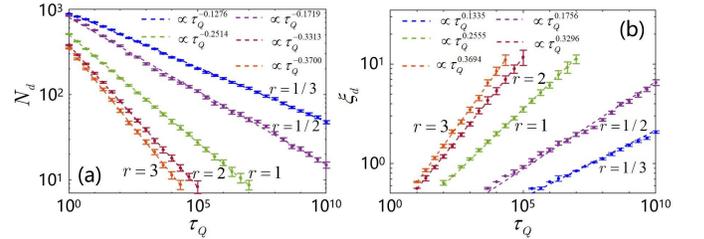}
\caption{(Color online)  Spatially universal scaling of the mean number of domains $N_{d}$ and the length of domain size $\xi_{d}$ for different quench times $\tau_{Q}$ and various power coefficients $r$. The number of domains is counted at the end of the evolution when the domains become stable. Error bars correspond to the standard deviation of $100$ runs. The parameters are the same as those used for Fig. \ref{Fig_Wave_Func_1}. }
\label{Fig_Domain_Corel_tauq_1}
\end{figure}

It is well known that the order parameter becomes nonzero near the critical point in the equilibrium process.
For the non-equilibrium dynamics in our system, the order parameter $\delta_{J}$ delays increasing near $+\hat{t}$.
For small $\tau_{Q}$, $\Delta J$ continues to delay until it increases, as shown by the dashed purple line in Fig. \ref{Fig_Order_Para_1}(d), while for larger $\tau_{Q}$, the delay of $\Delta J$ becomes small
(see the dotted blue line in Fig. \ref{Fig_Order_Para_1}(d)).
It is obvious to expect that the delay of $\Delta J$ vanishes when $\tau_{Q}\to\infty$; in this case, the non-equilibrium dynamics return to the equilibrium process. To study the relationship between the phase transition delay $b_d$ and quench time, we defined $b_d$ as
\begin{equation}\label{phase_delay_1}
{b_d}\sim \left|  \epsilon \right|\sim\left| {{g_{22}}\left( {\hat t} \right) - g_{22}^c} \right|\sim \left(\frac{r}{\tau _Q}\right)^{ - \frac{r}{{1 + rvz}}}.
\end{equation}
In Fig.~\ref{Fig_Delay1_vs_Tauq}, we show the universal scaling of the phase transition delay $b_{d}$ with respect to the quench time $\tau_{Q}$, and the numerical scaling for different $r$ is in good agreement with the analytical results.
To qualitatively study the universal scaling between the domain wall and the quench time. We count the number of domains $N_{d}$ by identifying the number of zero crossings of $J\left(x\right)$ at the end of the evolution. The number of domains for different $\tau_{Q}$ and various power coefficients $r$ are shown in Fig.\ref{Fig_Domain_Corel_tauq_1}(a). We observe that the mean number of domains follows a power-law scaling from the KZ mechanism, and we have
\begin{equation}\label{Domian_Scaling_Eq}
{N_d} \simeq {\hat \xi }^{-d}\sim\tau _Q^{ - \frac{{drv}}{{1 + rvz}}},
\end{equation}
where ${\hat \xi}$ is the correlation length at the freezing time. In Fig.\ref{Fig_Domain_Corel_tauq_1}(a), we clearly show a power-law scaling between the number of domains and the quench time, and the numerical scaling for various $r$ is in good agreement with the theoretical results.

\begin{figure}[htbp]
\centering
\includegraphics[scale=0.42]{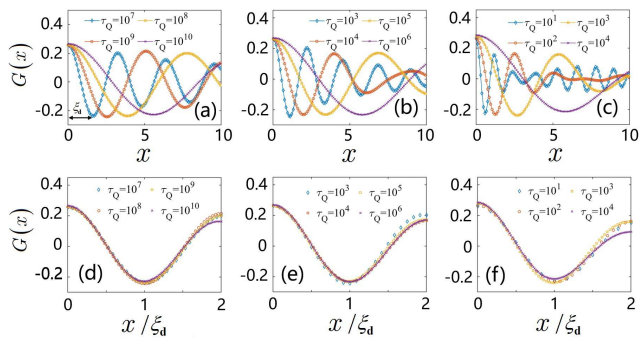}
\caption{(Color online) Spatial correlation functions $G(x)$ and its collapse for different $r$. (a,b,c) show
a typical example of the spatial correlation functions $G(x)$ for different quench times and power coefficients. Figs. (d, e, f) show that the correlation functions $G(x)$ collapse to a single curve when the distance is rescaled by the domain size $\xi_{d}$. }
\label{Fig_Correlation_length_1}
\end{figure}

To characterize the domain distribution, we then analyze the density-density correlation function,
\begin{equation}\label{Den-Correl_func}
G\left( x \right) = \int {J\left( x^{'} \right)J\left( {x^{'} + x} \right)d{x^{'}}}.
\end{equation}
We find that the correlation functions $G(x)$ for different quench times $\tau_{Q}$ all fall on a universal
curve when the distance is rescaled by the domain size $\xi_{d}$, as shown in the
Fig.\ref{Fig_Correlation_length_1}. A very peculiar feature of the correlation function is the appearance of oscillations. From
the correlation function, we extract the domain size $\xi_{d}$ or, equivalently, the distance between neighboring
domains. Our numerical results show that both the mean number of the domains and the domain size readily follow Kibble-Zurek scaling. \\
Combining the scaling of the phase transition delay $b_{d}$ and the domain wall $N_{d}$(or domain size $\xi_{d}$), we finally present the numerical critical exponents of the two-component BECs for different $r$ in Table \ref{Critical_Exponent_Table}.

\begin{table}
\normalsize
\begin{center}
\begin{tabular}{|l|l|l|l|l|}
\hline
  Power coefficient $r $ & 1/3 & 1/2 & 2/3 & 3/4  \\ \hline

Critical exponent $z$ & 2.001 & 1.957 & 1.964 & 2.013  \\ \hline

Critical exponent $v$ & 0.514 & 0.518 & 0.523 & 0.524  \\ \hline

  Power coefficient $r $  & 1 & 1.5 & 2 & 3  \\ \hline

  Critical exponent $z$ & 2.024 & 2.043& 2.034 & 2.041 \\ \hline

 Critical exponent $v$ & 0.512 & 0.509 & 0.508 & 0.503 \\ \hline
\end{tabular}
\caption{The numerical critical exponents($z$,$v$) of the two-component BEC for different $r$,
which are consistent with those of the analytical cases ($z$=2,$v$=1/2). }
\label{Critical_Exponent_Table}
\end{center}
\end{table}

\section{Summary and Discussions}

In summary, we have investigated universal non-equilibrium dynamics and defect formations in two-component BECs in nonlinear quenching across the miscible-immiscible phase transition.
By analyzing the Bogoliubov excitation, we obtain a power-law scaling relation among the defect density, the phase transition delay and the quench time.
Our work considers various quenching pathways with different power coefficients across the critical point.
We clearly show the formation of the domain defects for different quench times and power coefficients.
For large power coefficients $r$, the parameters of the system vary slowly, and the number of domain defects rapidly decreases, which indicates accessibility in experiments and the possibility for adiabaticity in finite-size systems.
The nonlinear quench approach provides a dynamic route that suppresses defect excitations in the fields of adiabatic state preparation and adiabatic quantum computation.
In addition, we quantify the temporal and spatial information of the domain defects using the spatial fluctuations of the local spin polarization and correlation length.
By combining the power-law relation between the number of defects, the phase transition delay and the quench time, we confirm that the critical exponents do not sensitively depend on the nonlinear quenches
of the phase transition.
Our study provides a dynamic path toward the suppression of nonadiabatic excitations.

\begin{acknowledgments}
This work was supported by GuangDong Basic and Applied Basic Research Foundation $($Grant No. 2021A1515111015$)$,  the National Natural Science Foundation of China $($Grant Nos. 11905032, 11904051, 11874112, 12274077$)$, the Natural Science Foundation of Guang-dong Province $($Grant No. 2021A1515010214$)$, the Key Research Projects of General Colleges in Guangdong Province $($Grant No. 2019KZDXM001$)$,  the Special Funds for the Cultivation of Guangdong College Students' Scientific and Technological Innovation $($Grant Nos. pdjh2022a0538, pdjh2021b0529$)$, and the Research Fund of Guangdong-Hong Kong-Macao Joint Laboratory for Intelligent Micro-Nano Optoelectronic Technology $($Grant No. 2020B1212030010$)$.
\end{acknowledgments}

%\bibliography{Non_KZM_Ref}% Produces the bibliography via BibTeX.

\end{document}